\documentstyle[referee]{mn}
\newcommand{\reference}{\bibitem}

\def\beq{\begin{equation}}
\def\eeq{\end{equation}}
\def\bey{\begin{eqnarray}}
\def\eey{\end{eqnarray}}
\def\beqarray{\begin{eqnarray}}
\def\eeqarray{\end{eqnarray}}

\def\Mpc{\,{\rm {Mpc}}}
\def\kpc{\,{\rm {kpc}}}

\def\kms{\,{\rm {km\, s^{-1}}}}
\def\msun{\rm M_\odot}
\def\vcir{V_{\rm c}}
\def\v200{V_{200}}

\def\my{\,{\rm M_\odot yr^{-1}}}
\def\E51{\,{\rm E_{51}}}
\def\nh{n_{\rm h}}
\def\S13{S_{-13}}
\def\fsigma{f_{\Sigma}}

\def\mug{\mu_{\rm g}}
\def\Mev{\dot {M}_{\rm ev}}
\def\pc{\,{\rm pc}}

\def\al{a_{\rm l}}
\def\au{a_{\rm u}}
\def\Mg{M_{\rm g}}
\def\ncl{n_{\rm cl}}
\def\fc{f_{\rm c}}
\def\Mps{M_{\rm ps}}
\def\rcr{r_{\rm cr}}
\def\mucr{\mu_{\rm cr}}

\input epsf

\title[]
{An analytic model for the galactic winds and mass outflows}
\author[]
{Chenggang Shu$^{1,2,4}$, H.J. Mo$^{2}$, Shude Mao$^{3}$
\thanks {E-mail: cgshu@center.shao.ac.cn,
hom@mpa-garching.mpg.de, smao@jb.man.ac.uk} \\
$^1$ Shanghai Astronomical Observatory, Chinese Academy of
Sciences, Shanghai 200030, China\\
$^2$ Max-Planck-Institut f\"ur Astrophysik,
      Karl-Schwarzschild-Strasse 1, Postfach 1317
      D-85741 Garching, Germany \\
$^3$ Jodrell Bank Observatory, Univ. of Manchester,
        Macclesfield, Cheshire SK11 9DL, UK\\
$^4$ The Joint Lab of Optical Astronomy, Chinese Academy of
Sciences}

\date{Accepted ........
      Received .......;
      in original form ......}
\pubyear{2002}
\begin{document}
\maketitle
\begin{abstract}
Galactic winds and mass outflows are observed both in nearby
starburst galaxies and in high-redshift star-forming
galaxies. In this paper we develop a simple
analytic model to understand the observed superwind phenomenon.
Our model is built upon the model of McKee \& Ostriker (1977)
for the interstellar medium. 
It allows one to predict 
how  properties of a superwind, such as wind velocity and mass
outflow rate, are related to  properties of
its star-forming host galaxy, such as size, gas density
and star formation rate. The model predicts a threshold
of star formation rate density for the generation
of observable galactic winds. Galaxies with more
concentrated star formation produce superwinds with higher
velocities. The predicted mass outflow rates are comparable
to (or slightly larger than) the corresponding star formation
rates.
We apply our model to both local starburst galaxies and high-redshift
Lyman break galaxies, and find its predictions
to be in good agreement with current observations. Our model
is simple, and so can be easily incorporated into numerical
simulations and semi-analytical models of galaxy formation.
\end{abstract}

\section {Introduction}

Galactic-scale bulk motions of gas, such as galactic winds and
mass outflows, have been observed both in local starburst
galaxies (Forbes et
al 2000; Heckman et al 2000 and references therein) and in high
redshift galaxies (Pettini et al 2000, 2001; Dawson et al 2002;
Adelberger et al 2002).
Evidence for such superwinds and mass outflows comes from X-ray
emission produced by the hot gas or optical line emission from the
warm gas associated with star-forming galaxies
(Martin, Kobulnicky \& Heckman 2002;  Strickland et al
2002). Based on medium-resolution spectra of Na D
($\lambda\lambda 5890, 5896$) and X-ray observations
of some local starburst galaxies, Heckman et al (2000)
constrained wind velocities in the range $\sim 400-800\kms$
and mass outflow rates comparable to
star formation rates (hereafter SFRs).
Using observations of nebular absorption lines and Ly-$\alpha$
emission lines, Pettini et al (2000) and Adelberger et al (2002)
showed that galactic winds are common in high-redshift
Lyman break galaxies. The inferred wind velocities for these
objects range from several hundred $\kms$ to about 1000$\kms$,
with a median value of about $500\kms$. For the
bright gravitationally lensed Lyman-break galaxy, MS 1512-cB52
($z=2.72$), Pettini et al (2000, 2001) estimated
the mass outflow rate to be $\sim 60\my$, comparable to the star
formation rate of this system derived from its UV luminosity.

These bulk motions of gas (superwinds) in star-forming galaxies
are believed to be driven by the
kinetic energy from supernova (hereafter SN)
explosions, and may
change the thermal and chemical properties of the intergalactic
medium (IGM) that forms galaxies (Ferrara, Pettini \& Schekinov 2000;
Furlanetto \& Loeb 2002).
It is therefore essential to have a proper understanding
of these phenomena in order to understand galaxy formation
itself.

So far, our understanding of galactic superwinds is quite
incomplete. Based on the SN remnant evolution model of McKee \& Ostriker
(1977), Efstathiou (2000) established a multiphase interstellar
medium (ISM) model in which galactic winds result from the
expansion of hot phase gas. Combined with other physical
prescriptions such as the infall of the cooling gas and star
formation, he investigated the formation and evolution 
of disk galaxies in this model, and studied the dependence 
of mass outflow from a galaxy on the circular velocity of 
its host halo. 
Silk (2001, 2002) proposed an 
analytic model for the feedback process and outflow
in galaxies 
based on disk gravitational instability and multiphase ISM. 
Numerically there have also been attempts to use simulations
to understand how galactic winds can be produced
(e.g. Tomisaka \& Bregman
1993;  Suchkov et al 1994; Mac Low \& Ferrara 1999).
However, current simulations still lack the resolution
to deal with processes such as the evolution of
supernova remnants and the formation of multi-phase gas
that may be essential for the formation of superwinds.
Because of this, simple assumptions are usually made about
superwinds when studying their impact on the IGM
(e.g. Strickland \& Stevens 2000; Scannapieco, Ferrara \&
 Broadhurst 2000; Scannapieco, Ferrara \& Madau 2002;
Theuns et al 2002; White, Hernquist \& Springel 2002).

In this paper, we attempt to construct a simple analytic model for
superwinds. Our model is based on the model of McKee \& Ostriker
(1977) for supernova evolution in the ISM. Our goal is to
understand how properties of superwinds, such as wind velocity and
mass outflow rate, are related to the properties of star-forming
galaxies, such as size, gas density and SFR. The establishment of
such relations will allow one to include superwinds in simulations
and semi-analytical models of galaxy formation. We test our model
by comparing its predictions with observations of superwinds in
local starburst galaxies and in high-redshift Lyman-break
galaxies.

Our basic model
for the generation of superwinds is similar 
to that of Efstathiou (2000), as both his paper and
ours are based on McKee \& Ostriker (1977). 
While Efstathiou focused on the details about the 
assembly and star formation in two example galaxies 
(one dwarf-type and the other Milky-Way type), 
we focus on the statistical properties of superwinds 
in star forming galaxies based on simple empirical 
models about star formation.  
In addition, we compare our theoretical predictions 
with observations for local starburst galaxies
and high-redshift Lyman-break galaxies.

\section {Model}

In this section, we will use the model of McKee \& Ostriker (1977,
hereafter MO77) for supernova evolution in the interstellar medium
to construct a model of galactic winds and mass outflows for star
forming galaxies. Before going into the details, we summarize here
the basic idea in our modeling (see also 
Efstathiou 2000, hereafter E2000, for more details).

In the model of MO77, cold star-forming gas in the ISM is assumed
to be in the form of cold clouds surrounded by the warm ISM with
temperature of about $ \la 10^{4}\rm K$. Massive stars evolve and
explode in a few million years as supernovae (SNe), and SN
remnants propagate to form a low-density hot medium with a
temperature of about $10^{5-6}\rm K$. The expansion of the SN
remnants compresses and sweeps out material from the cold clouds
in the shock front to produce large-scale bulk motions in the ISM,
which can be accelerated by the pressure in the SN remnants. We
assume such bulk motion to be responsible for the observed
superwinds and mass outflows. In such a scenario, the properties
of the superwind from a galaxy are expected to depend not only on
the properties of the ISM, but also on other properties of the
galaxy in consideration. In what follows, we quantify the main
processes involved in our model.

\subsection{Supernovae evolution, interstellar media and galactic winds}

Star formation generally takes place in giant molecular clouds
which consist of many small and dense clouds.  The differential
number density distribution of clouds as a function of cloud
radius is assumed to follow a power-law (MO77, E2000)
\beq
\frac{dN_c}{da} = n_0 a^{-4},~~~ \al \leq a \leq \au,
\eeq
where
$\al$ and $\au$ are lower and upper limits for the cloud radii,
and $n_0$ is a normalization constant. We take, following MO77,
\beq
\al =0.5\pc, ~~~ \au/\al = 20.
\eeq
We caution that the exact
value of the lower limit radius ($\al$) is unclear. 
It is generally believed to be in the range of 0.5 to $1\pc$ (MO77); 
recent observations by Olmi \& Testi (2002) suggested
that $\al \sim 0.55\pc$.

It is easy to show by integration that the local cloud number
density, $n_{\rm cl}$, is related to $n_0$ by
\beq \ncl =
n_0/3\al^3, ~{\rm for~~} \au \gg \al.
\eeq
Assuming that $\bar
\rho_{\rm c}$ is the mean mass density within individual clouds,
the mean cold gas density $\bar \rho_{\rm cold}$ can be written as
\beq
\bar \rho_{\rm cold} = \int_{\al}^{\au}
\frac{dN_c}{da}\frac{4\pi}{3}a^3 \bar \rho_{\rm c} da =
\frac{4\pi}{3}\bar \rho_{\rm c} n_0 {\rm ln}\frac{\au}{\al}.
\eeq
Thus, the volume filling factor of the cold gas is
\beq
\fc =
\frac{\bar \rho_{\rm cold}}{\bar \rho_{\rm c}}.
\eeq
The porosity
$P$ is related to $\fc$ by $\fc \equiv e^{-P}$.

According to MO77, an expanding SN remnant will sweep up the
ambient cold gas with a total mass of 
\beq 
M_{\rm ev} =540\E51^{6/5} \nh^{-4/5} \Sigma^{-3/5} \msun, 
\eeq 
where $\E51$ is
the energy output by one SN in unit of $10^{51}\rm erg$, $n_{\rm
h}$ is the hot gas density interior to the SN remnant and $\Sigma$
is the evaporation parameter 
\beq 
\Sigma = \frac{\gamma}{4\pi \al
\ncl \phi_k} = \frac {\gamma \al^2 {\rm ln}(\au / \al)}{\phi_k}
\frac {\bar \rho_{\rm c}}{\bar \rho_{\rm cold}}. 
\eeq 
Here
$\gamma$ is the ratio of the blast wave velocity to the isothermal
sound speed of the hot phase, which is usually taken to be 2.5
(E2000), and the parameter $\phi_k$ denotes the efficiency of
conduction relative to the classic thermal conductivity of the
clouds. $\phi_k$ is roughly in the range of 0.1 to 0.01 due to the
suppression of heat conduction by tangled magnetic fields,
turbulence, etc (E2000).

As argued by several authors (Silk 1997, 2001, 2002; E2000; Clarke
\& Oey 2002), star formation in galaxies may proceed in such a way
that the porosity $P$ is always maintained as a constant value close to
unity in star formation regions. This constancy in the porosity
$P$ is supported by observations in the Milky Way, and can be
physically understood as follows. If the porosity were too large,
the cold gas fraction would be small and the star formation
activity would be reduced. The
hot gas will then cool quickly, and hence leading to a lower
porosity. Conversely, if the porosity were too small, the cold gas fraction
would be large. As the dynamical timescale
is likely to be short in the star formation regions, star
formation will ensue and hence
lead to a higher SFR and higher porosity. In
the present study, we assume the porosity to be unity everywhere
in a star formation region. 

Adopting $\au /\al = 20$, we can rewrite the evaporation parameter
($\Sigma$) as 
\beq  \label{eq:Sigma} 
\Sigma = 752
\fc^{-1}\left(\frac {\gamma}{2.5}\right ) \left (\frac
{\phi_k}{0.01}\right )^{-1} \left( \frac {\al}{\pc}\right)^2
\pc^2. 
\eeq 
Notice that the value of $\Sigma$ is not very
sensitive to the ratio $\au/\al$ because it enters eq.
(\ref{eq:Sigma}) logarithmically. For convenience, we define a new
parameter 
\beq \label{eq:fsigma} 
\fsigma =
\frac{\Sigma}{\Sigma_{\odot}}, ~~~~ \Sigma_{\odot} = 95\pc^2 
\eeq
which is the evaporation parameter normalized to the value close
to the solar neighborhood (E2000). Numerically, we find
\beq 
\fsigma = 21.5 \left( \frac{\fc}{e^{-1}}\right
)^{-1}\left(\frac {\gamma}{2.5}\right ) \left (\frac
{\phi_k}{0.01}\right )^{-1} \left( \frac {\al}{\pc}\right)^2. 
\eeq

Following MO77 and E2000, we can also obtain the temperature of
the hot gas 
\beq \label{eq:th} 
T_{\rm h} =
6.6\times10^5(\S13\E51\fsigma/\gamma)^{0.29}\, {\rm K}, 
\eeq 
and the mass evaporation rate per unit volume 
\beq \label{eq:rhoev}
\dot \rho_{\rm ev} = 2.7 \times 10^{-10} \S13^{0.71} \gamma^{0.29}
\E51^{0.71} \fsigma^{-0.29} \msun\pc^{-3}\rm yr^{-1}, 
\eeq
respectively, where $\S13$ is defined as the the SN explosion rate
in unit of $10^{-13}\rm pc^{-3}yr^{-1}$. From equations
(\ref{eq:th}), (\ref{eq:rhoev}) and the definition of $\fsigma$
[eq. (\ref{eq:fsigma})], we see that the temperature $T_{\rm h}$
and the mass evaporation  rate $\dot \rho_{\rm ev}$ are in fact
independent of the parameter $\gamma$. Note that ${\dot \rho}_{\rm
ev}$ increases with supernova rate density (which is proportional
to the SFR density) as a power of $0.71$. Thus, for a given total
star formation, the total outflow rate increases with the volume
of the star formation region.

The hot phase will first expand to form super bubbles which
compress the ambient ISM. The compressed ambient ISM will be
driven to form a ``wind'' with a velocity comparable to the
isothermal sound speed of the hot phase, which is given by 
\beq
C_i = (kT_{\rm h}/\mu m_{\rm p})^{1/2} = 37T_5^{1/2} \kms 
\eeq
with  $T_5$ the hot gas temperature in unit of $10^5 \rm K$ and
$\mu$ the mean molecular weight per particle; we take $\mu=0.61$,
a value appropriate for a fully ionized primordial gas.  Because
of the conservation of the specific enthalpy, the wind will
accelerate and reach a bulk terminal speed to form a superwind
(E2000). The
terminal wind velocity is related to the isothermal sound speed by
\beq \label{eq:vwind} 
v_{\rm wind} = \Gamma_w C_i. 
\eeq 
We take
$\Gamma_w \approx \sqrt{2.5}$, instead of a value $\sqrt{5}$; our
value is appropriate when part of the thermal energy is lost
radiatively [see Appendix B in E2000 for details].
Observationally, only when the initial sound speed of the hot gas
is larger than $\sim 100 \kms$, can galactic winds be
readily observed (e.g. Heckman et al 2000). From eq.
(\ref{eq:vwind}), this sound speed corresponds to roughly
$160\kms$ in the terminal velocity. 

Note that dark matter halos, which dominate the potential wells of
galaxies, have not entered our discussions explicitly. Thus, the
local properties of outflows, such as wind velocity and
evaporation rate per unit volume, depend only on the properties of
the ISM in the star-forming region. This is
consistent with the observations of X-rays from hot gas in
galaxies and the detailed analysis of wind velocities by Heckman
et al (2000), who found that terminal velocities of the superwinds
are almost independent of the galaxy potential wells. However, the
dark halo of a galaxy will play an important role in determining
whether the outflow can eventually escape from the galaxy or fall
back into the galaxy.

It is clear from equations (\ref{eq:th}) and (\ref{eq:rhoev})
that the model predictions
for the wind properties (velocity and mass outflow rate) depend on the
choices of the values of
$\phi_k$ and $\al$. As discussed above, the values of these two
parameters are probably in the ranges
$0.01\la \phi_k \la 0.1$ and  $0.5\pc \la \al \la 1\pc$.
Using equations (\ref{eq:th}) and (\ref{eq:rhoev}),
one can estimate that these ranges correspond to
a factor of $\la 2$ in the predicted wind
velocity, and a factor of $\la 3$ in the predicted mass
outflow rate. These factors should serve as an indication
of the uncertainties in our model predictions.

One assumption we made is that the porosity $P$ retains close to
unity. While this seems to be supported by observation but hard to
prove rigorously (Silk 2001), it is important to examine the
dependence of our results on the value of $P$. In general,
galactic winds are more efficient during the early adiabatic stage
of SN remnant evolution, when the porosity $P \la 1$. In this
case, SN remnants do not overlap with each other.  
The physical prescriptions we developed for $P=1$ are valid
for $P<1$, except that the wind velocity $v_{\rm wind}$ and mass
outflow rate $\dot \rho_{\rm ev}$ must be modified by a factor of
$P^{-1/7}$ and
$(1-e^{-P})/[P^{5/7}(1-e^{-1})]$, respectively (MO77).
For a range of $P$ from 0.2 to 2, the wind velocity changes by
less than 25\% while the mass outflow rate changes by less than
20\% from our canonical results (for $P=1$). Clearly both $v_{\rm
wind}$ and $\dot \rho_{\rm ev}$ are insensitive to $P$. As SN
remnants evolve, they may overlap with each other 
and $P$ may becomes larger than 1. In this case, the model 
will break down. As discussed above, in real galaxies, 
$P$ may be regulated so that its value cannot be 
much larger than 1. We therefore believe that the assumption 
$P=1$ does not lead to significant error in our results.

\subsection{Star Formation in Spherical Regions}

So far we have only considered superwinds and mass outflows
launched locally from a particular location in a star formation
region. The derived superwinds and mass outflows therefore only
depend on the local star formation activities. The observed wind
velocity and mass outflow rate from a galaxy should be a proper
average and sum of the wind velocity and outflow rate over the
whole star formation region. It is these global quantities that
can be directly compared with observation. In this subsection, we
consider a simple model where star formation takes place uniformly
in a spherical region. 
Here the wind and outflow are simply evaluated by the SFR density and the
integration over the volume within star forming 
region. We treat individual star forming cells as
source of winds but ignore the detailed transfer
of energy and mass from the center to the outer region.
Our treatment is obviously simplistic; a more
careful consideration must consider wind interactions which
result in the eventual outflow from the surface of the 
spherical region. 

For a spherical region of uniform star formation,
$\S13$ can be written as
\beq
\S13 = \frac{\dot M_*}{\Mps V} = 2\times 10^3\left(\frac{\dot
M_*}{100\my}\right)
{\left(\frac{\Mps}{125\msun}\right)}^{-1}
\left(\frac{R}{1\kpc}\right)^{-3} ,
\eeq
where $\dot M_*$ is the
SFR within radius $R$ and $\Mps$ is the stellar mass of
star formation per supernova explosion. For a Salpeter mass function
($n(M) dM \propto M^{-2.35} dM$), $\Mps \approx 125
\msun$ with a stellar mass range from 0.1 to $50\msun$.

Using equation (\ref{eq:vwind}), we can estimate the galactic wind
terminal velocity
\beq  \label{eq:sphericalWind}
v_{\rm wind} =
623\kms\left[\left(\frac{\dot M_*}{100\my}\right)^{0.29}
\left(\frac{\Mps}{125\msun}\right)^{-0.29} \left(\frac
{R}{1\kpc}\right)^{-0.87}
{\cal K}^{0.29} \right]^{1/2}
\eeq
and the corresponding total mass evaporate rate
\beq \label{eq:sphericalMev}
\Mev =
133 \my \left ( \frac {\dot M_*}{100\my} \right)^{0.71}
\left(\frac{\Mps}{125\msun}\right)^{-0.71}
\left(\frac {R}{1\kpc}\right)^{0.87}
{\cal K}^{-0.29},
\eeq
For convenience we have defined $\cal K$
as a function of $\fsigma$ and $\gamma$ as
\beq \label{eq:K}
{\cal K} = \left(\frac{\fsigma}{21.5}\right)
 \left(\frac{\gamma}{2.5}\right)^{-1}\,,
\eeq
which gives the dependence of $v_{\rm wind}$ and $\Mev$ on the
properties of the ISM.

For fixed ${\cal K}$, the wind velocity
depends only on the SFR density,
${\dot M}_*/ R^3$, because the temperature of the hot gas,
hence the wind velocity of the driven ISM, depends only on the SN explosion
rate density [cf. eq. (\ref{eq:rhoev})].
The wind velocity increases with the
SFR density.
Thus, galaxies with a more compact cold gas distribution
will produce winds with higher velocity, no matter what
the total amount of cold gas and total SFR are.
This implies that superwinds are more likely to be observed
in local starbursts and high-redshift star forming galaxies where the 
gas density is high and star formation activity is
concentrated.  For a given SFR density, the
wind velocity increases with the lower limit of the cloud radii,
because larger clouds leads to a larger value of
$\fsigma$ [cf. eq.(\ref{eq:fsigma})]. The wind velocity also increases
with decreasing $\phi_k$, because a higher $\phi_k$ implies
a higher energy loss due to thermal transfer.
As discussed in the last subsection, we adopt the lower limit of the wind
velocity to be
$\sim 160\kms$ which, according to Heckman (2000),
is the lower limit that can lead to an observable superwind.
This limit directly translates into a lower limit on the
SFR density, $0.01-0.1\my\kpc^{-3}$,
below which no observable wind is expected.

On the other hand, the mass outflow rate has
an additional dependence on the size of star formation
region.
In Fig. 1, we show the predicted mass
outflow rate  as a function of SFR for three sets of
parameters. The solid, dashed and dotted
lines denote the results for $(\phi_k, \al)$ equal to ($0.1, 0.5\pc$),
($0.01, 0.5\pc$), and $(0.01, 1\pc)$, respectively.
The thin and thick lines show results for the two adopted
size of star formation region, $0.1\kpc$ and $1\kpc$,
respectively. For a given total SFR,
a larger star formation region gives a larger mass outflow
rate, because the increase of the outflow rate density
with SFR density is slower than linear
(see eq. \ref{eq:rhoev}).
Note, however, this increase of
$\Mev$ with size cannot go indefinitely, because our
assumption that $P=1$ will fail when the gas density
becomes so low that the volume is too large for the SN remnants to
fill.

\begin{figure} \label{fig:Mevfunc}
\epsfysize=9.5cm
\centerline{\epsfbox{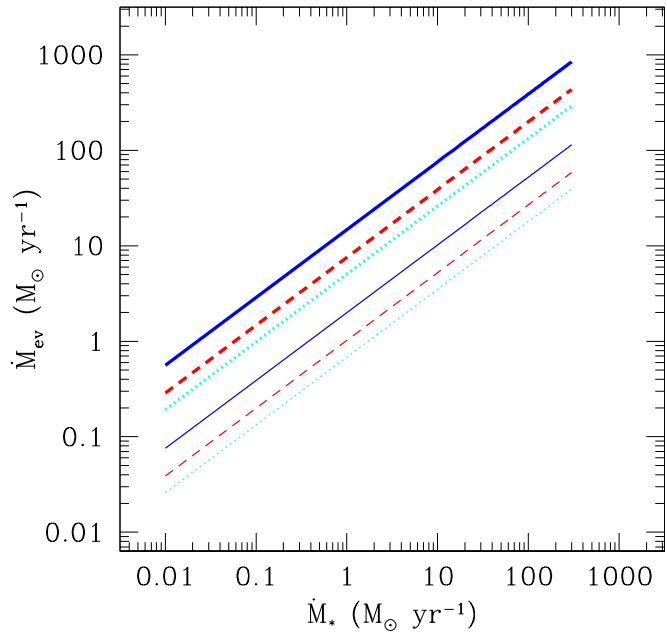}}
\caption{The predicted mass outflow rate as a function of the
SFR  for three sets of parameters, shown as the solid,
dashed and dotted lines corresponding to  $(\phi_k, \al)$ equal to
($0.1, 0.5\pc$), ($0.01, 0.5\pc$), and $(0.01, 1\pc)$,
respectively. The thin and thick lines are for two different sizes of
star formation regions, $0.1\kpc$ and $1\kpc$, respectively.}
\end{figure}

 Since the total power in the wind, which is proportional
to ${\dot M}_{\rm ev} v_{\rm wind}^2$, should be proportional to the total
SFR, the dependence of ${\dot M}_{\rm ev}$
on model parameters can be understood in the same way as
the dependence of wind velocity on model parameters.  From
Fig. 1 we see that the predicted mass outflow rates
are comparable to (or higher than) the corresponding star
formation rates for a wide range of model parameters. This is consistent
with observation, e.g. Heckman et al. (2000)
and Pettini et al. (2000; 2001; 2002).

\subsection{Star Formation in Galactic Disks}

In this subsection, we consider another case where
star formation is assumed to be in thin galactic disks.
This may be more realistic than the spherical model
described above, because the cold gas supporting star formation
generally has angular momentum and is found in
a rotation-supported thin disk.

For simplicity, we assume the velocity dispersion
($\sigma_g$) of the cold clouds in a disk to be
 constant everywhere. Throughout this paper,
we choose $\sigma_g$ to be $10\kms$, a value consistent with
observation (Stark \& Brand 1989). The
equation of hydrostatic equilibrium gives the following solution of
cold gas
density distribution in the vertical direction:
\beq
\rho(z) = \frac {\mug}{2H_g}{\rm sech^2}\left (\frac
{z}{H_g}\right),
\eeq
where $\mug$ is the cold gas surface density and $H_g$ is the scale height
given by
\beq
H_g = \frac {\sigma_g^2}{\pi G \mug} = 74\left(
\frac{\mug}{\msun\pc^{-2}}\right )^{-1} \left(
\frac{\sigma_g}{\kms}\right)^2 \pc,
\eeq
and $G$ is the gravitational constant (Spitzer 1942).

Kennicutt (1998) studied star formation in a wide range
of physical conditions, ranging from quiescent galactic disks
to starburst regions. He derived an empirical law
for the SFR per unit area as a function of the cold gas
surface density,
\beq \label{eq:kennicutt}
\dot \mu_* = 2.5\times 10^{-10}\left( \frac{\mug}{\msun\pc^{-2}}\right
)^{1.4} \my \pc^{-2}.
\eeq
It then follows that  the SFR density is
\beq
\dot \rho_* = \frac {\dot \mu_*}{2H_g} = 1.7 \times 10^{-12}\left(
\frac{\mug}{\msun\pc^{-2}}\right )^{2.4}
\left(\frac{\sigma_g}{\kms}\right)^{-2} \my \pc^{-3},
\eeq
and the SN explosion rate $\S13$  is
\beq
\S13 = 10^{13} \times \frac{\dot \rho_*}{\Mps} = 1.35\times 10^{-1}
\left(\frac{\mug}{\msun\pc^{-2}}\right )^{2.4}
\left(\frac{\sigma_g}{\kms}\right)^{-2}
\left(\frac{\Mps}{125\msun}\right)^{-1},
\eeq
where $\Mps$ is again the mass of
stars formation corresponding to one supernova explosion.

Substituting the above equation into eqs. (\ref{eq:th}) and (\ref{eq:vwind}),
we can infer the wind velocity
$v_{\rm wind}$ to be
\beq \label{eq:diskvwind}
v_{\rm wind} = 150\kms
\left(\frac{\mug}{\msun\pc^{-2}}\right )^{0.35}
\left(\frac{\sigma_g}{\kms}\right)^{-0.29}
\left(\frac{\Mps}{125\msun}\right)^{-0.15}
{\cal K}^{0.15}.
\eeq
The corresponding mass outflow rate per unit area
can be obtained by multiplying
the mass  outflow rate per unit volume and the scale height
\beq \label{eq:diskMuev}
\dot \mu_{\rm ev} \approx 2 H_g \dot{\rho}_{\rm ev} =
 5.1 \times10 ^{-9} \my \pc^{-2}
\left(\frac{\mug}{\msun\pc^{-2}}\right )^{0.7}
\left(\frac{\sigma_g}{\kms}\right)^{0.58}
\left(\frac{\Mps}{125\msun}\right)^{-0.71}
{\cal K}^{-0.29},
\eeq
where $\cal K$ is defined in eq. (\ref{eq:K}).

 As one can see, for given $\sigma_g$ and ${\cal K}$,
the wind velocity increases with $\mu_{\rm g}$.
Thus, superwinds are more likely to occur in compact
systems of cold gas. The wind velocity decreases
with $\sigma_g$, because a higher value of $\sigma_g$
implies a larger disk thickness, and hence a lower density
of cold gas. The mass outflow rate per unit area
also increases with gas surface density, because of
the dependence of the SFR per unit area
on the surface density of cold gas.

As for the spherical star formation region in last
subsection, we adopt the same threshold for the wind velocity,
below which winds may not yield observable signatures
(Heckman et  al. 2000). This threshold corresponds to a lower limit
on the SFR per unit area, which can be obtained from
eq. (\ref{eq:diskvwind}). For the reasonable ranges of $\phi_k$
and $\al$, i.e., $\phi_k=0.01-0.1$ and $\al=0.5-1\pc$,
the limit  is about
\beq \label{eq:threshold}
{\rm SFR}_{\rm th} \sim 0.01 -  0.05 \my \kpc^{-2},
\eeq
which, from eq. (\ref{eq:kennicutt}),
corresponds to a cold gas surface density about $10 - 40\msun\pc^{-2}$.
This limit is consistent with the observational estimation quoted by
Heckman (2001).

\begin{figure}
\epsfysize=9.5cm
\centerline{\epsfbox{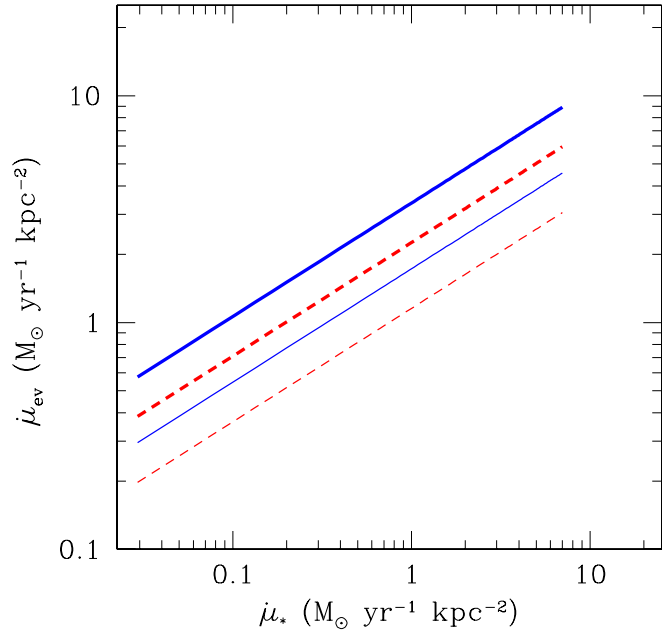}}
\caption{The predicted mass outflow rate per unit area as a function
of the SFR per unit area. The solid and dashed lines are
the results
where the lower limit of the cloud radius is chosen to be $0.5\pc$
and $1\pc$, respectively. The thin and thick lines denote results
with the parameter $\phi_k$ chosen to be 0.01 and 0.1,
respectively.}
\end{figure}

The predicted
mass outflow rate per unit area as a function of the star formation
rate per unit area is shown in Fig. 2.  The solid and dashed lines are
results when the lower limit of the cloud radius is chosen to be $0.5\pc$
and $1\pc$, respectively. The thin and thick lines denote results
with the parameter $\phi_k$ chosen to be 0.01 and 0.1,
respectively. For any given
$\phi_k$ and $\al$, the mass outflow
rate per unit area increases with the SFR
per unit area, as pointed out before.
Note that the predicted mass outflow rates per unit area are comparable to
(or higher than) the SFRs per unit area,
which is consistent with observation (Heckman et al 2000; Pettini et
al 2000; 2001; 2002).

 So far we have considered only local properties of
winds in galactic disks. To facilitate direct
comparison between model predictions and observations,
we must consider globally-averaged quantities of the wind.
As a simple example, we assume that the cold gas is distributed in an
exponential disk
\beq
\mug = \mu_0 {\rm exp} (-r/r_d), ~\mu_0 = {\Mg \over 2 \pi r_d^2}
\eeq
where $\mu_0$ is the central surface density of cold gas,
$\Mg$ is the total cold gas mass and $r_d$ is
the scale length for the disk. Note that
our results are not sensitive to the assumed cold
gas surface density distribution because the cold gas surface
density enters the wind velocity and mass outflow rate
only with moderate power-law indices (0.35 and 0.7)
in eqs. (\ref{eq:diskvwind}) and (\ref{eq:diskMuev}),
respectively.

As discussed above, there is a threshold in
the SFR per unit area (eq. \ref{eq:threshold}), i.e.,
the lower limit of the cold gas surface density,
above which observable galactic winds can be produced. If the
central surface density $\mu_0$ for a disk galaxy is lower than this,
no observable galactic wind will occur. When
$\mu_0$ is above the threshold, the threshold implies a critical
radius $\rcr$ for any given ${\cal K}$, beyond which no outflow
will occur. The critical radius satisfies
\begin{equation} \label{eq:mucr}
\mucr=\mu_0 {\rm exp} (-\rcr/r_d).
\end{equation}
It is easy to calculate the total mass outflow rate
within this critical radius
\begin{eqnarray} \label{eq:diskMev}
\dot M_{\rm ev} &=& 6.2\times 10^{-2}\my \left(\frac {r_d}{\kpc}\right)^2
\left(\frac{\mu_0}{\msun\pc^{-2}}\right)^{0.7}\nonumber\\
&&\times \left(\frac{\sigma_g}{\kms}\right)^{0.58}
\left(\frac{\Mps}{125\msun}\right)^{-0.71}
{\cal K}^{-0.29}{\cal F}(0.7, \mu_0)\nonumber \\
&=&9.8\my \left(\frac {\Mg}{10^9\msun}\right)
\left(\frac{\mu_0}{\msun\pc^{-2}}\right)^{-0.3}\nonumber\\
&&\times
\left(\frac{\sigma_g}{\kms}\right)^{0.58}
\left(\frac{\Mps}{125\msun}\right)^{-0.71}{\cal K}^{-0.29}
{\cal F}(0.7, \mu_0)\,,
\end{eqnarray}
where the function ${\cal F}(x, \mu_0)$ is defined as
\beq \label{eq:calF}
{\cal F}(x, \mu_0) = 1-\left[1-x {\rm ln}\left(\frac{\mucr}
{\mu_o}\right)\right]
\left(\frac{\mucr}{\mu_o}\right)^{x}.
\eeq
${\cal F}(x, \mu_0)$ describes the fraction of area
that contributes to the observable galactic wind.
For any given $x$, ${\cal F}$ will first increase with
$\mu_0$ for small values of  $\mu_0$, and
then decrease with increasing $\mu_0$ for large
values of $\mu_0$.

The observed superwind velocity  is an appropriate
 average of the wind velocities over the whole disk.
Since the mass outflow rate per unit area is proportional
to $\mu_{\rm g}^{0.7}$, we define the global wind velocity
to be the average of the wind velocities weighted by
$\mu_{\rm g}^{0.7}$. This wind velocity is
\beq \label{eq:diskVwind1}
V_{\rm wind} = 68 \kms
\left(\frac{\mu_0}{\msun\pc^{-2}}\right )^{0.35}
\left(\frac{\sigma_g}{\kms}\right)^{-0.29}
\left(\frac{\Mps}{125\msun}\right)^{-0.15}
\left(\frac {{\cal F}(1.05,\mu_0)}{{\cal F}(0.7,\mu_0)}\right)
{\cal K}^{0.15}.
\eeq
In reality, the observed wind velocity depends on where
the wind originates, and our weighting scheme can only
serve as an approximation. We have also made calculations
using a weight proportional to $\mu_{\rm g}$ and to
$\mu_{\rm g}^{1.4}$ (i.e. to the surface density of SFR). The results
for these three weighting schemes
differ only by 20\%.
Note that for given ${\cal K}$ the wind velocity for an exponential disk
depends only on the central surface density of cold gas,
while the total mass outflow rate for the disk depends in
addition on the scale length (or the total cold gas mass).

\begin{figure}
\epsfysize=9.5cm
\centerline{\epsfbox{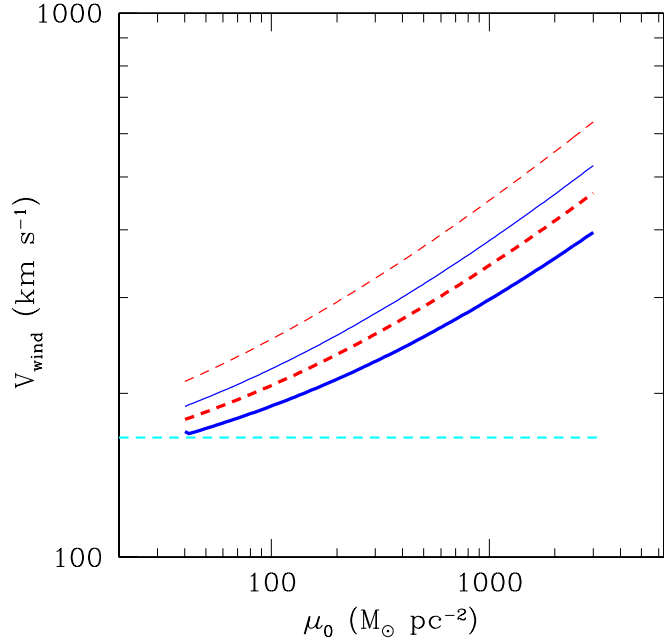}}
\caption{The predicted mass outflow rate weighted wind velocity as a
function of $\mu_0$ where the solid and dashed
lines have the same notations as
that in Fig. 2
with the dotted horizon line marking the lower limit of the
velocity, $\sim 160\kms$. Here the velocity dispersion of the cold gas
clouds is assumed to be $10\kms$. }
\end{figure}

Fig. 3 shows the the predicted global wind velocity for an exponential
disk with $\mu_0 \ga \mucr$ as a function of cold gas
central surface density $\mu_0$ for four different combinations
of $\phi_k$ and $\al$.
The notation of the figure is the same as that in Fig. 2
with the dotted horizon line marking the lower limit of the
velocity, $\sim 160\kms$, as discussed above.
For given $\phi_k$ and $\al$,
the predicted global wind velocities
increase with the cold gas surface density $\mu_0$.
Galaxies with low cold gas surface density (e.g.,
low surface brightness galaxies or high-surface brightness
galaxies with low cold gas surface density
such as our Milky Way)  will not
produce observable galactic winds  because their star
formation rate is too low.

\begin{figure}
\epsfysize=9.5cm
\centerline{\epsfbox{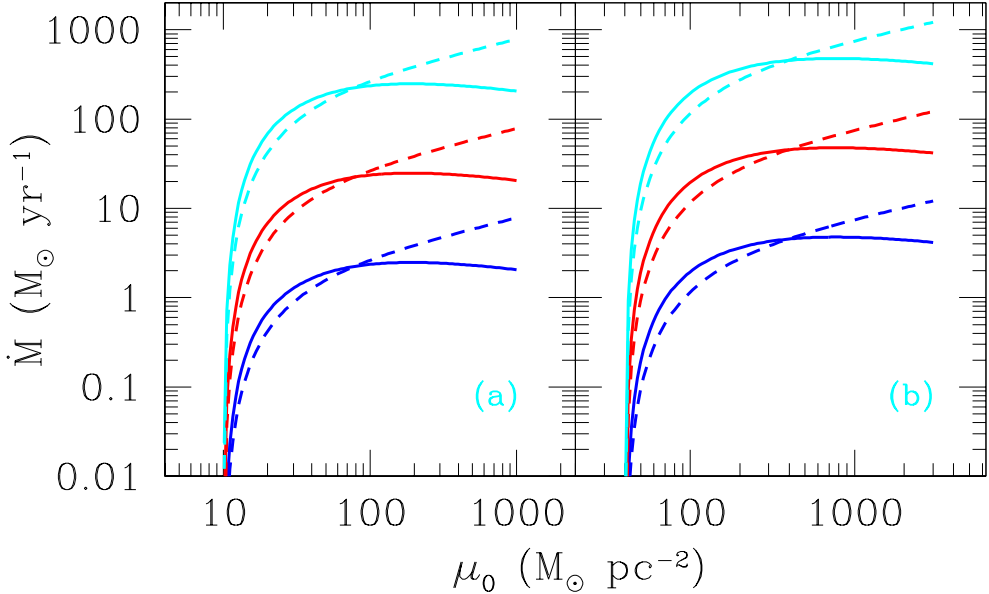}}
\caption{The predicted mass outflow rate (solid lines)
and the corresponding star
formation rate (dashed lines) as a function of $\mu_0$ for given the total
cold gas mass, where lines from top to bottom denote the total
cold gas mass to be $2\times 10^{11}\msun$, $2\times 10^{10}\msun$ and
$2\times 10^{9}\msun$ respectively.
Here the velocity dispersion of the cold gas
clouds is assumed to be $10\kms$, and (a) and (b) denote the results
of ($\phi_k$, $\al$)
equal to (0.01, $1\pc$) and (0.1, $0.5\pc$), respectively.}
\end{figure}

As an example, Fig. 4 plots the predicted total mass outflow rate
(solid lines)
and the corresponding SFR (dashed lines) as a function of
$\mu_0$ with (a) and (b) denoting the results of ($\phi_k$, $\al$)
equal to (0.01,
$1\pc$) and (0.1, $0.5\pc$), respectively.
The lines from top to bottom are results for a total
cold gas mass to be $2\times 10^{11}\msun$, $2\times 10^{10}\msun$ and
$2\times 10^{9}\msun$, respectively. From the figure we see that
the cold gas central density $\mu_0$ must be larger than a critical
value $\mucr$ (about 10 and $40\msun\pc^{-2}$ for the values of
$\phi_k$ and $\al$ adopted here respectively) for an observable wind to
occur. If $\mu_0 < \mucr$, no observable
galactic wind will be produced no matter how large the total
cold gas mass is.

For $\mu_0 \ga \mucr$, the total mass outflow rates increase with
the total mass of cold gas for a given $\mu_0$ because of the
increase of the SFR and the star formation region
which can produce observable outflows.
For a given total cold gas mass, the
predicted mass outflow rate first increases rapidly with $\mu_0$,
because the gas mass that can produce observable
winds [i.e. $r_{\rm cr}$ in eq. (\ref{eq:mucr})] increases with $\mu_0$.
The mass outflow rate as a function of $\mu_0$ reaches a plateau
and then decreases when $\mu_0$ is larger than
$\sim 10^{3}\msun\pc^{-2}$. This happens
because when $\mu_0$ is high enough, the total gas mass
that can produce observable galactic winds is saturated
while the SFR density increases with increasing $\mu_0$.
This increase in SFR density reduces the mass outflow rate,
again because the increase of ${\dot\rho}_{\rm ev}$
with SFR density is slower than linear (see eq. \ref{eq:rhoev}).
Note that the predicted mass outflow rates
are comparable to the SFR for a very wide range
of $\mu_0$, which is consistent with current observations.

\section{Comparisons with observations}

As discussed in the introduction,
many observational studies have investigated galactic winds
both in local starburst galaxies and in high-redshift star
forming galaxies. In this section, we examine whether
the predictions of our model can match the current
observational data.

For definiteness, we will adopt $\phi_k = 0.01$ and $\al = 1\pc$
throughout this section. As we discussed in Sec. 2.1, other
plausible choices of these parameters may result in
a lower wind velocity by a factor of $\la 2$ and
a higher mass outflow rate by a factor of $\la 3$.
The choice is quite arbitrary, since the exact values
of $\phi_k $ and $\al$ appropriate for star forming
galaxies are not known {\it a priori}. Our chosen values do give
reasonable agreement with the observational results 
to be discussed below.

\subsection{Local starburst galaxies}

Based on the Na D absorption lines, Heckman et al (2000) estimated
 galactic wind velocities for local starburst
galaxies. They found that the wind velocities are in the range of 400 to
800$\kms$. They also estimated mass outflow rates, which they
claimed to be comparable to the SFRs.
The observational results for the mass
outflow rates are only qualitative and we have shown that
they are consistent with our model predictions.
In what follows, we will primarily focus on the wind velocities
which are more accurately determined.

At first we assume that the star formation takes place
within spherical regions. We can then estimate the global galactic winds
for these galaxies using equation (\ref{eq:sphericalWind}), provided
that their SFRs and sizes are known.
We estimate the SFRs for these starburst
galaxies using their FIR luminosities (Kennicutt 1998)
\beq \label{eq:FIRsfr}
\frac {\rm SFR}{\my} = \frac {L_{\rm FIR}}{5.8\times 10^{9} L_{\odot}}.
\eeq
The sizes of the star forming regions for the galaxies in
consideration are not available. As a
rough estimate, we take the sizes to be the same as the
observational slit widths, which, according to Heckman et al (2000),
is a reasonable match to the typical 
sizes of powerful starbursts.
The observational data are from Table 4 in Heckman et al
(2000). The predicted wind velocities
vs the observed ones are plotted as crosses in Fig. 5(a). The lines in
the figure
denote the agreement
between observation and model predictions. As can be seen,
our model predictions roughly match the observations within the model and
observational uncertainties.

\begin{figure}
\epsfysize=9.5cm
\centerline{\epsfbox{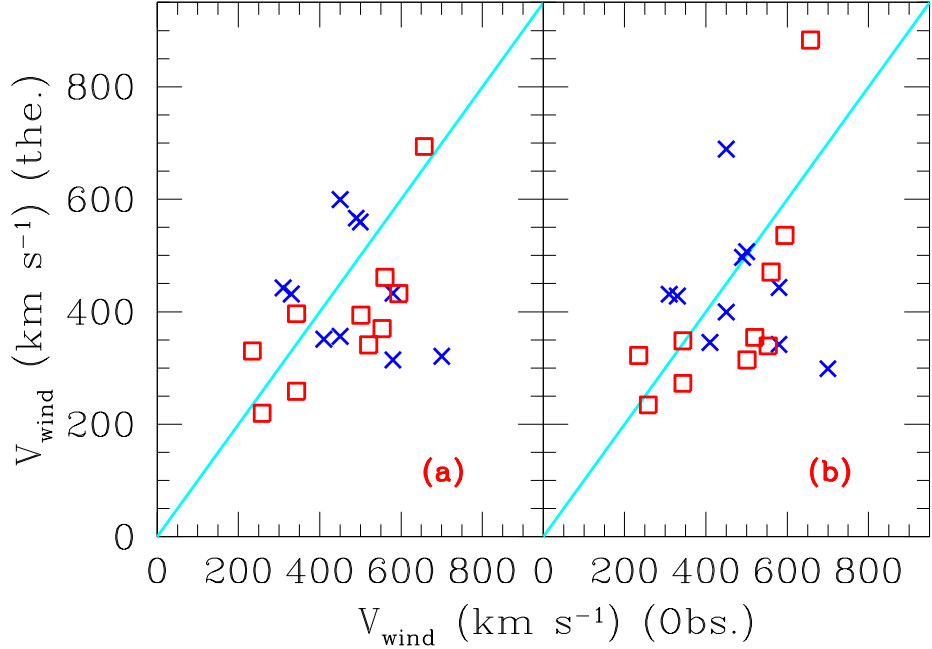}}
\caption{The predicted wind velocities vs observed values for
local starburst galaxies. The crosses denote the
observations of Na D absorption lines and
open squares denote the estimated results based on the the hot gas
temperatures from ROSAT and ASCA. The lines denote the agreement
between observation and model predictions. (a) for spherical
star formation region;
(b) for exponential star formation disk (see the text for details).}
\end{figure}

Another way to empirically estimate the super wind velocities for
galaxies is based on X-ray observations. Galactic
winds will spread out the hot gas within galactic halos and contribute
to the soft component of extended X-ray emission. As the hot gas temperature
can be inferred from X-ray observations (e.g., from ROSAT or ASCA),
we can estimate the wind velocities from the temperature of
the soft X-ray components using equation (\ref{eq:th}).
The galaxies we selected are similar to those in Heckman et al
(2000). They are NGC4449 (Della Ceca, Griffiths \& Heckman 1997),
NGC2146 (Della ceca et al. 1999), NGC253, NGC3079, M82, NGC4631
(Dahlem, Weaver \& Heckman 1998),
NGC1569 (Della ceca et al. 1996), Arp299 (Heckman et al 1999), NGC6240
(Iwasawa \& Comastri 1998), NGC1808 (Awaki et al. 1996).
Note that we assume the relation between wind velocity
and soft X-ray temperature to be $V_{\rm wind} = \sqrt{2.5k T_X}$
rather than $\sqrt{5kT_X}$, the relation adopted by Heckman et al (2000).
The reason for our choice is discussed below eq. (\ref{eq:vwind}).

We can estimate the SFRs of these galaxies based
on their FIR luminosities according to equation (\ref{eq:FIRsfr}).
The sizes of their star formation regions are adopted from the
corresponding ${\rm H}\alpha$ observations, which can be found in the
references listed above. The predicted wind velocities
can then be obtained using eq. (\ref{eq:sphericalWind}). The open
squares in Fig. 5(a) indicate the predicted
wind velocities and the estimated values from the X-ray observations.
We find that the predictions  match the estimated values reasonably well.

As pointed out in Sec. 2.3, the predicted results are not sensitive
to the assumed gas surface distribution if we assume
starbursts take place in exponential disks.
Assuming that the cold gas
in these galaxies is distributed exponentially and that observed  star
formation region  contain just one half of the total star formation
activity in the galaxies, we can also predict the superwind
velocities for the above starburst galaxies based on equation
(\ref{eq:diskVwind1}). The model predictions versus observational
results are shown in Fig. 5(b). Here again,
the model predictions match  observation.

\subsection{Lyman break galaxies}

The UV dropout method has been very successful in identifying active star
forming galaxies (Lyman break galaxies, hereafter LBGs) at a
redshift of $z \approx 3$ (Steidel, Pettini \& Hamilton 1995).
Much observational and theoretical work has
investigated their physical nature (see Mo \& Fukugita 1996; Mo, Mao \&
White 1999; Kauffmann et al. 1999; Katz, Hernquist \& Weinberg 1999;
Shu 2000; Shu, Mao \& Mo 2001). Based on observations of Lyman alpha
 and nebular emission lines, LBGs are also inferred
to display  galactic winds (Pettini et al 2001; 2002; Adelberger et al
2002). LBGs display a wide range of wind
velocities,  250 to nearly $1000\kms$ with a
median value about 500-600$\kms$. As for local starburst galaxies,
the mass outflow rates for LBGs are difficult to establish observationally.
Pettini et al (2000; 2002) measured the mass outflow rate
for a specific LBG, the lensed and magnified MS 1512-cB58; the value
is $\sim 60\my$, comparable to its SFR
obtained from its UV luminosity.

\begin{figure}
\epsfysize=9.5cm
\centerline{\epsfbox{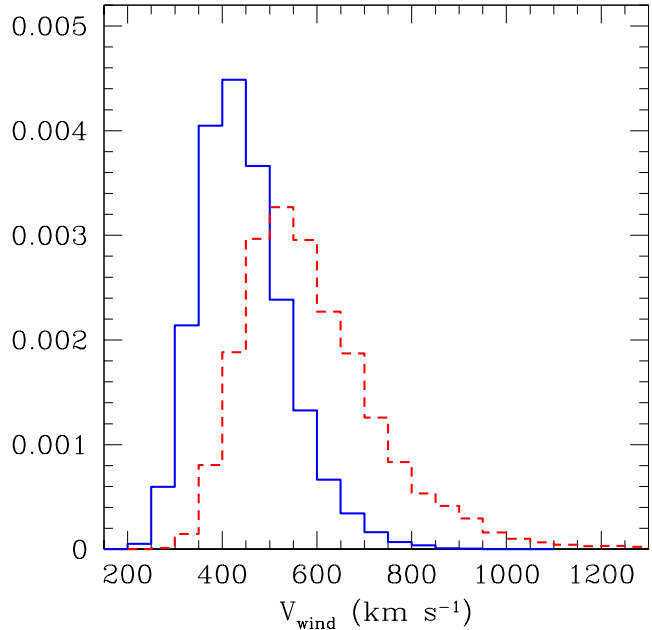}}
\caption{The predicted distribution of the galactic winds for LBGs.
The solid and dashed histograms denote results where
star formation is assumed to take place in
spherical regions and thin exponential disks, respectively.}
\end{figure}

At present it is unclear whether the star formation in
LBGs occurs in quasi-spherical regions or more in a disk-like
geometry. For this reason, we will consider both
scenarios below in turn.

If the star formation activities in LBGs are quasi-spherical and
homogeneous, then we can use equations equations (\ref{eq:sphericalWind})
and (\ref{eq:sphericalMev}) to  estimate the mass outflow rates and
galactic wind velocities for LBGs if we know the star formation
rates and the sizes of their star formation regions.
Following the procedure outlined in Shu, Mao \& Mao (2001), we
adopt the SFR distribution of LBGs from their
dust-corrected UV luminosity and a log-normal size distribution.
We use  Monte Carlo simulations to select galaxies following
these SFR and size distributions. We then
substitute these two quantities into eqs.
(\ref{eq:sphericalWind}-\ref{eq:sphericalMev}) to obtain
the predicted distributions of the galactic wind velocities and the
corresponding mass outflow rates. The results are shown as the solid
histograms in Figures
6 and 7,
respectively. The predicted winds have velocities in the range
from 200 to 900$\kms$ and a median value of 450$\kms$,
in agreement with the quite limited observational results. The
predicted median value of the mass outflow rate is about 80$\my$,
similar to the inferred value for MS 1512-cB58.

\begin{figure}
\epsfysize=9.5cm
\centerline{\epsfbox{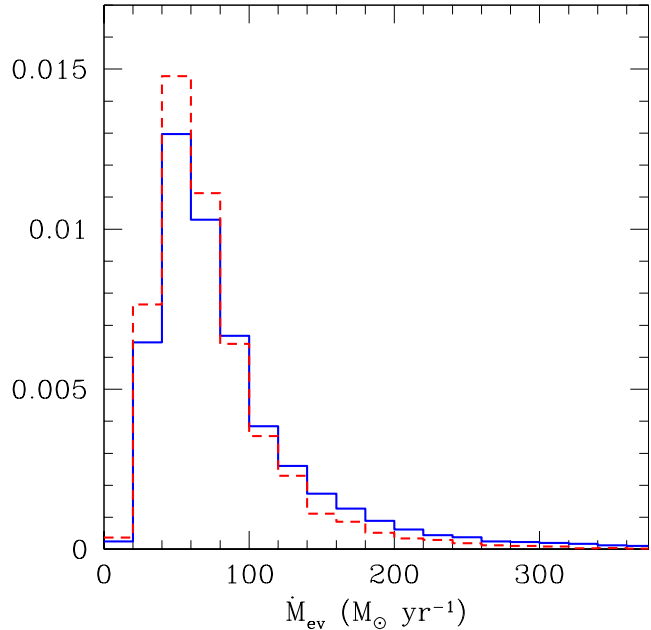}}
\caption{The predicted distribution of the mass outflow rates for LBGs.
The solid and dashed histograms denote results where star formation
is assumed to occur in spherical regions and thin exponential
disks, respectively.}
\end{figure}

The second scenario we consider for LBGs is that their star formation
occurs in a pre-assembled disk. For simplicity, we assume that
the cold gas surface density has an exponential distribution.
In this case, the model discussed in Sec. 2.3 can be applied
to the LBG population. We use a procedure identical to that
used in the spherical model to simulate the LBG population.

Based on eqs. (\ref{eq:diskMev}-\ref{eq:diskVwind1}),
the estimated outflow rates and wind velocities can be obtained.
The dashed histograms in Figs. 6 and 7 show the model predictions for the
distributions of the galactic wind velocities and the corresponding mass
outflow rates for LBGs, respectively.
It can be seen that the median value of the predicted
wind velocities for LBGs is $\sim 600\kms$. The median value
of the predicted mass outflow rates is about
$60\my$. Both are consistent with observations.

\section{Escaping galactic winds as a function of
$V_{\rm c}$ and redshift}

As we discussed in Sec. 2, galaxies with more compact cold gas
distributions (and 
hence more compact star formation activity) will produce
stronger galactic winds. Galactic winds will thus occur preferentially in
local starburst galaxies and high redshift
galaxies. If the speed of a galactic wind is smaller than the escape
velocity of the host galaxy, the mass outflow will fall
back into the galaxy and form a galactic fountain. Otherwise, the
outflow will escape from the host. This we call an
``escaping'' outflow. The mechanical energy
of the outflow can heat the the intergalactic medium (IGM) while heavy
elements contained in the wind 
can chemically enrich this gas. It is therefore
important and interesting to estimate the fraction of galaxies (by number)
that will have escaping outflows at different redshifts. We
address this question in the currently preferred $\Lambda$CDM cosmogony
with a matter density $\Omega_m=0.3$, a cosmological
constant $\Omega_{\lambda}=0.7$, a Hubble constant $h =
70\kms\Mpc^{-1}$; the power-spectrum is described by a shape
parameter $\Gamma = 0.2$ and the usual normalisation
constant $\sigma_8 = 0.9$.

\begin{figure}
\epsfysize=9.5cm
\centerline{\epsfbox{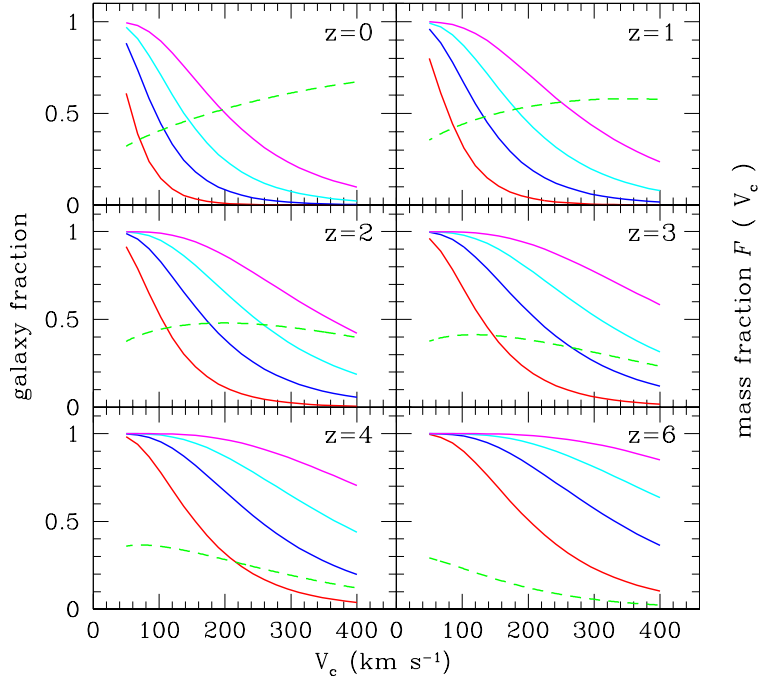}}
\caption{The predicted fraction of galaxies where the superwinds
escape from their dark matter haloes is shown as a function
of circular velocity at six different redshifts. In each panel,
the four solid lines correspond to a gas faction
of $m_{\rm g}=$0.1, 0.05, 0.025
and 0.01 (from top to bottom), respectively. The predicted mass fraction
$F(\vcir)$ in halos for a given
$\vcir$ is shown as a dashed line at each redshift in
the current $\Lambda$CDM cosmogony.}
\end{figure}

We assume that the distribution of cold gas within a galaxy is
exponential; our results will not change significantly if we adopt other
profiles (see Sec. 2.3). We use Mo, Mao \& White (1998) to model
the formation of disks in dark matter haloes. In this model, the
disk properties are determined by the halo circular
velocity, $\vcir$, a dimensionless spin parameter,
$\lambda$, and the fraction of total mass ($m_{\rm g}$)
that settles into the exponential disk. The spin parameter
follows a log-normal distribution  with a
median ${\bar \lambda}=0.05$ and a dispersion $\sigma_{\rm ln
\lambda}=0.5$ (e.g., Warren et al 1992; Lemson \& Kauffmann 1999).
Once these three parameters ($\vcir, \lambda$, and $m_{\rm g}$) 
are specified, the disk properties
and hence the star formation and outflow properties can
be evaluated using the formalism developed here. In Fig. 8,
we show the number fraction of galaxies with escaping winds occuring
in the disks 
as a function of circular velocity at six different redshifts.
For each redshift, four solid lines are plotted, corresponding to
four different $m_{\rm g}$ values. Note that, for these results,
we have averaged over the spin parameter distributions.
In the same figure, we also show the mass fraction
$F(\vcir)$ in halos with a given circular velocity 
$\vcir$ in logrithmic bin at different redshifts, i.e., 
\beq
F(\vcir) d {\rm \log \vcir} = \frac {d F(>\vcir, z)} {d \rm \log
\vcir} d {\rm \log \vcir} , 
\eeq  
where $F(>\vcir,z)$ is the mass fraction in haloes with circular
velocity larger than $\vcir$
as predicted by the updated Press-Schechter formalism (c.f. Mo \& White
2002). From the dashed line, one can easily estimate the mass fraction
of total halos and the
typical halo circular velocity which will produce escaping
mass outflows.

As expected, for any given $m_{\rm g}$, a 
 larger fraction of small halos produces escaping outflows at
any redshift because of their shallower potential wells. Hence,
during their evolution, a significant fraction of their baryons will
be lost due to the galactic winds. It also implies that
most small halos in the local universe are dark matter dominated
which is consistent with observation (e.g. Persic, Salucci \& Stel 1996).
Given $\vcir$, galaxies with larger $m_{\rm g}$ have a larger
probability to produce an escaping outflow at any redshift
because of their more active star formation. As the redshift increases,
more and more galaxies produce escaping outflows because of the
increase in their SFR densities. For example at $z=3$, more than 80\% galaxies
with $\vcir \la 200\kms$  will produce 
mass outflows that escape for $m_{\rm g}\ga 0.05$.
Hence escaping
outflows will occur commonly in LBGs because the median value of
their circular velocity is about $150\kms$ and  their
$m_{\rm g}$ may be $\ga 0.06$ (Shu, Mao \& Mo 2001).
Hence, a significant number of galaxies will contribute to
the heating of the IGM or ICM. Subsequent galaxy formation therefore
will occur in preheated intergalactic media, as argued by
Mo \& Mao (2001). Galactic winds therefore should be taken into
account for high-redshift galaxies.

As a comparison, it is interesting to study how the star formation
rate is partitioned in haloes with different circular velocities.
Specifically, we study the differential probability distribution
of star formation rate in logrithmic scales of $\vcir$, i.e.,
\beq
f(\vcir)d {\rm \log \vcir} = \frac {\rm SFR (\vcir)}{\int\rm SFR
(\vcir)d {\rm \log \vcir}} d {\rm \log \vcir},
\eeq
where SFR is the star formation rate in galaxies for a given circular velocity
$\vcir$ averaged over all spin parameter distributions.
Note that this distribution function is normalised to unity
for $V_c$ between 30~km/s and 400~km/s. 

We define the accumulative probability distribution function for galaxies with 
escaping outflows, $f_{\rm ro}(<\vcir)$, as a function of $\vcir$,
\beq
f_{\rm ro}( < \vcir) = \frac {\int_{30}^{\vcir} \rm SFR_{esc} (\vcir) d
{\vcir}}{\int_{30}^{400}\rm SFR (\vcir)d {\vcir}}, 
\eeq 
where $\rm SFR_{esc}$ is the star formation rate
in galaxies with escaping outflows for a given
circular velocity $\vcir$. 

To obtain these two probability distributions,  we have assumed that all
galaxies are disk galaxies. Furthermore, the gas fraction $m_{\rm g}$ in
the disks is taken to be the fraction of gas that eventually cool
in their host haloes; we take the cooling function
from Sutherland \& Dopita (1993) assuming the metallicity of
0.01 solar value. The influence of the feedback on $m_{\rm g}$ is also
taken into account by adopting 
\beq
m_{\rm g} = \frac {m_{\rm g0}} { 1 + (\frac{150\kms}{\vcir})^2}
\eeq
as suggested by Dekel \& Silk (1986) and White \& Frenk (1991).
Here $m_{\rm g0}$ is
the maximum baryon fraction that can cool in the haloes to form disks.

\begin{figure}
\epsfysize=9.5cm
\centerline{\epsfbox{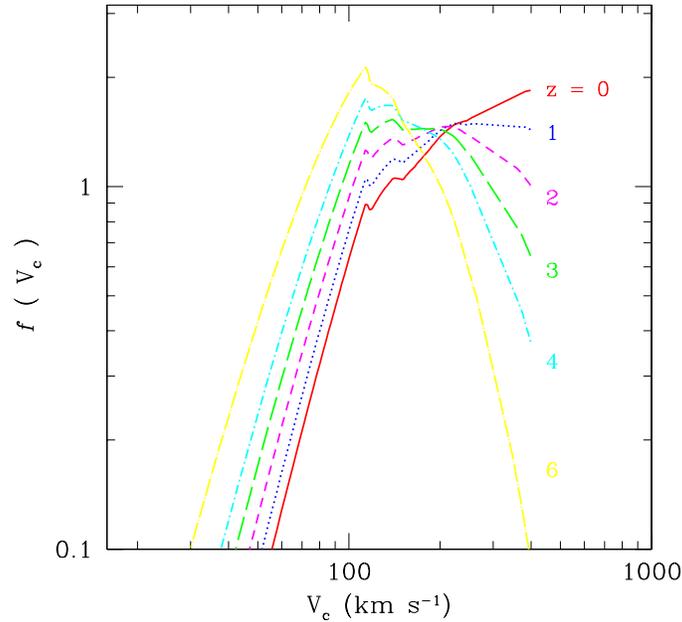}}
\caption{The predicted fraction of galaxies weighted by the SFR as a function
of circular velocity at six different redshifts. Gas
cooling and feedback effect have been taken into account and 
the currently favoured $\Lambda$CDM cosmogony is adopted (see text for
details).} 
\end{figure}

We show the two probability distributions 
in Figures 9 and 10 for six different redshifts,
respectively. We can see from Fig. 9 that the SFR fraction of galaxies
peaks at $\vcir \sim 100$ - $200\kms$  
at redshift $z > 1$. However, at lower
redshift, most of the SFR is contributed by  more
massive galaxies because the feedback at low redshift is not as
efficient and most baryons within massive halos can cool to form stars.
From Fig. 10, we see that the contribution of SFR for galaxies
with escaping outflows are dominated by galaxies with $\vcir \sim
200\kms$ at redshift $z\la 3$, while at higher $z$
significant contribution comes from systems with 
$\vcir \sim 150\kms$. 
This is because, although at lower redshift the SFR is shifted 
toward systems with higher $\vcir$, escaping
superwinds are more difficult to produce in these systems.
At $z> 3$, more than $40\%$ of all stars are formed in systems 
where escaping winds are expected.   
Note that a change in the value of $m_{\rm g0}$ does
not affect the result significantly. At high redshift, 
the number density of high $\vcir$ systems goes 
down rapidly and so not many stars can form 
in these systems.

\begin{figure}
\epsfysize=9.5cm
\centerline{\epsfbox{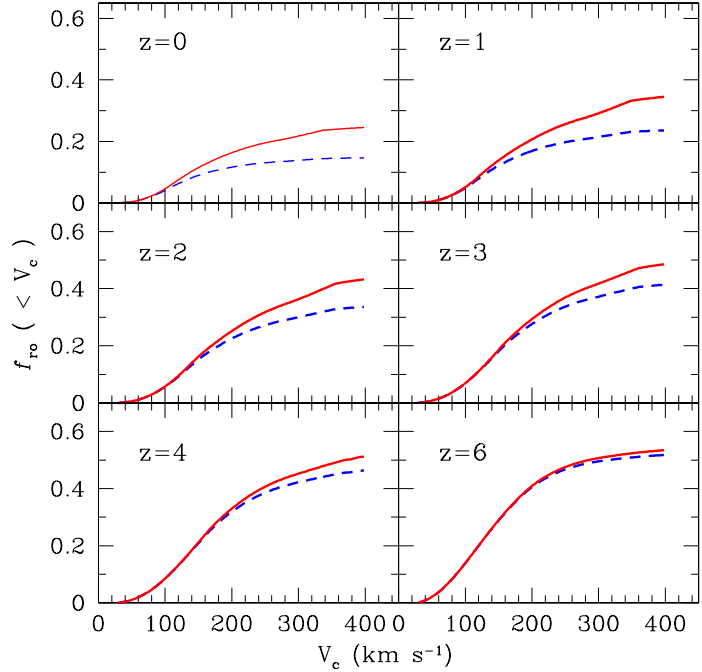}}
\caption{The predicted accumulative probability distribution function
of galaxies  
with escaping outflows as a function
of halo circular velocity at six different redshifts. For each redshift,
the solid and dashed
lines denote results for $m_{\rm g0}=$0.1 and  0.05, respectively. Gas cooling
and feedback effect have been taken into account and 
the currently favoured $\Lambda$CDM cosmogony is adopted (see text for
details).} 
\end{figure}

\section{Summary}

In this paper, we develop a simple analytic model
to understand superwinds and mass outflows that are observed
in local starburst galaxies and high-redshift star-forming
galaxies. Our scenario is based on the model of Mckee \& Ostriker (1977)
for the evolution of SN remnants in the ISM, which is similar to
Efstathiou (2000). We show that
the properties of superwinds, such as wind velocity and mass
outflow rate, depend not only on the properties of the ISM,
but also on the global properties of star forming galaxies.
Our main conclusions are:
\begin{itemize}
\item
Observable winds are produced only in systems
where the density of SFR is higher
than some threshold. This is consistent with the
observational results given by Heckman (2001) and
Heckman et al (2000), and implies that low
surface-brightness galaxies and high surface-brightness
galaxies with low cold gas surface density
cannot produce large-scale superwinds.
\item
The velocity of superwinds driven by SN explosions
depends only on the SFR density.
Galaxies in which current star formation is confined in
a more compact region can produce winds with higher velocity.
This implies that superwinds are expected in local starbursts
and in high-redshift star-forming galaxies.
The mass outflow rate depends, in addition, on the size of star
formation region; a larger star formation region allows
more mass to be loaded in the wind.
\item
The predicted mass outflow rates are comparable to
or higher than the corresponding SFRs,
consistent with current observations both for local starburst
galaxies (Heckman et al 2000) and for high redshift star-forming
galaxies (Pettini et al 2000; 2001; 2002).
\item
The predicted wind velocity and outflow rate
have no explicit dependence on the properties of dark
halos which dominate the potential wells.
Thus, galactic winds and mass outflows can occur
in a variety of halos, provided that the cold gas density
is high enough. This is in agreement with the observation
that there is no strong correlation between wind properties
and the total mass of a galaxy (Heckman et al 2000).
The potential well of a galaxy will, however, determine
whether the outflows  escape from the galaxy or return.
\item
The fraction of galaxies with superwinds that
will eventually escape from their dark matter haloes
is a function of circular velocity and redshift.
Our model predicts that the fraction is high
in low circular velocity systems such as dwarf galaxies
although their contribution to the total SFR is small.
More interestingly, we find that the fraction is
high for galaxies at high redshifts, such as the
LBGs at redshift $z=3$. The winds will undoubtedly heat
and chemically contaminate the IGM, and hence
have important implications for subsequent galaxy formations
in the preheated IGM.

\end{itemize}

We apply our model to make predictions for the properties
of the winds expected from local starburst galaxies and high-redshift
Lyman-break galaxies. These predictions can match many
of the observed properties. We therefore believe
that our model catches the main points required to  model the superwind
phenomenon. Our model is also simple, and so can be easily
incorporated into numerical simulations and semi-analytical
models of galaxy formation.

\section*{Acknowledgement}

We thank Simon White for carefully reading
the manuscript and useful suggestions. 
CS acknowledges the financial support of MPG for a visit to
MPA.
This project is partly supported by the Chinese National Natural
Foundation and the NKBRSF G1999075406.

\end{document}